%% file: main.tex
\documentclass{article}
\usepackage[utf8]{inputenc}
\usepackage[hyphens]{url}
\usepackage{hyperref} 
\usepackage{etoolbox}
\usepackage{amssymb}
\usepackage{amsmath}
\usepackage{clrscode3e}
\usepackage[nodayofweek,level]{datetime} 
\usepackage{enumitem} 
\usepackage{multirow} 
\usepackage[calc]{picture} 
\usepackage{color}
\usepackage{authblk}

\newcommand{\wan}{WebAuthn}

\newcommand{\cpp}{C\texttt{++}}

\AtBeginEnvironment{quote}{\itshape}
\title{Technical Report on a Virtual CTAP2 WebAuthn Authenticator}
\author[1]{Chris Culnane}
\author[2]{Christopher J. P. Newton}
\author[2]{Helen Treharne}
\affil[1]{Castellate Consulting Ltd, chris@castellate.com} 
\affil[2]{Department of Computer Science, University of Surrey, \{c.newton\}\{h.treharne\}@surrey.ac.uk}

\date{December 2020}

\usepackage{graphicx}

\begin{document}
\maketitle
\begin{abstract}
Even though passwordless authentication to online accounts offers greater security and protection from attack, passwords remain prevalent. Passwordless authentication adoption is impacted by the slow adoption of external hardware keys required to generate the security keys within the authentication protocol. We have developed a virtual WebAuthn authenticator in order to provide an extensible open source platform for understanding the associated standards of WebAuthn and CTAP2.  Our authenticator provides secure software authentication for devices that do not have access to a physical hardware interface. Our authenticator also provides an alternative to an external physical hardware key and supports the use of a trusted platform module (TPM) on a device to generate the security keys within a WebAuthn protocol.

\end{abstract}
\section{Introduction}
This technical report covers the development of a Virtual CTAP2 WebAuthn authenticator. The authenticator is intended to provide a platform for testing and development of WebAuthn/CTAP2 protocols and extensions. In addition to a software only authenticator it also provides a proof of concept implementation of a Trusted Platform Module (TPM)~\cite{tpmtcg} based authenticator, with associated interfaces and libraries for using a TPM as the underlying credential store. 

This document provides a high-level overview of the design and implementation. API documentation is provided in the \textit{docs} folder of the source code repository \url{https://github.com/UoS-SCCS/VirtualWebAuthn}.

This report starts by providing a brief overview of WebAuthn\cite{webauthn} and its history, followed by a short overview of the CTAP2\cite{ctap2} standard and how it relates to WebAuthn. Section \ref{sec:virtual_authenticator} provides an architecture and implementation description of the virtual authenticator we have implemented. This provides a detailed overview of the classes and how the framework has been developed.

The report was developed as part of an EPSRC project\footnote{https://gow.epsrc.ukri.org/NGBOViewGrant.aspx?GrantRef=EP/N028295/1} that focused on Data to Improve the Customer Experience (DICE). The project's main application domain was intelligent transport systems (ITS) but the scope included ensuring security and data privacy when using web services, for example in the case of smart ticketing~\cite{DBLP:journals/tifs/HanCSTWW20} and emerging technologies~\cite{10.1007/978-3-030-64455-0_2} that could be applicable in the ITS domain. In this report and that associated GitHub we make reference to the authenticator that we developed as a DICEkey. It has no particular significance other than to provide a link to the project.

\section{WebAuthn}
WebAuthn is a standard API that provides strong authentication services on the web. In essence it provides a public key credential that allows a user to log in to a web service without the use of a password. A number of cryptographic techniques are used to deploy such credentials in a way that is both safe and privacy preserving. 

There are at least two parties (Relying Party and User), and a number of components, which form the WebAuthn API. At the core there is a Relying Party (RP) that is seeking to use the WebAuthn API to create and use public key credentials that belong to users to authenticate them to their web service. By extension there is also a user, the individual who is authenticating themselves to the Relying Party. Bridging between these two parties are a number of components, whose control varies from the RP to the User and somewhere in between. 

The RP must implement the server-side component of the WebAuthn protocol, along with the necessary cryptographic functionality to facilitate the verification and evaluation of public key credentials (PublicKeyCredential). Each user will also have to have an authenticator, this can come in a number of forms, including hardware based tokens, trusted platform modules, and smartcards. Whilst those are the conventional types of authenticator, the only requirement is that it is a cryptographic entity that is capable of performing the necessary cryptographic operations. 


The bridge between a User controlled authenticator and an RP server-side component is provided by a client or User-Agent - typically a web browser, but not exclusively. As such, the authenticator does not communicate directly with the RP, instead the requests and responses are proxied through the client. 

The WebAuthn standard defines the roles, properties, and capabilities that the various components should have, and contents of the requests and responses that will be communicated between them. However, it does not define a concrete mechanism for how to interface with the authenticator. Instead it defines an abstract model of what an \textit{Authentiactor} is and leaves it up to implementers to define a concrete protocol. One such example is the FIDO2 CTAP2 (client-To-Authenticator-Protocol). CTAP2 defines a concrete authenticator that can be interfaced with via USB, Bluetooth, or NFC. As such CTAP2 can be considered to be a concrete implementation of the WebAuthn abstract model of an authenticator.

\begin{figure}[ht]
\begin{center}

\includegraphics[width=\textwidth]{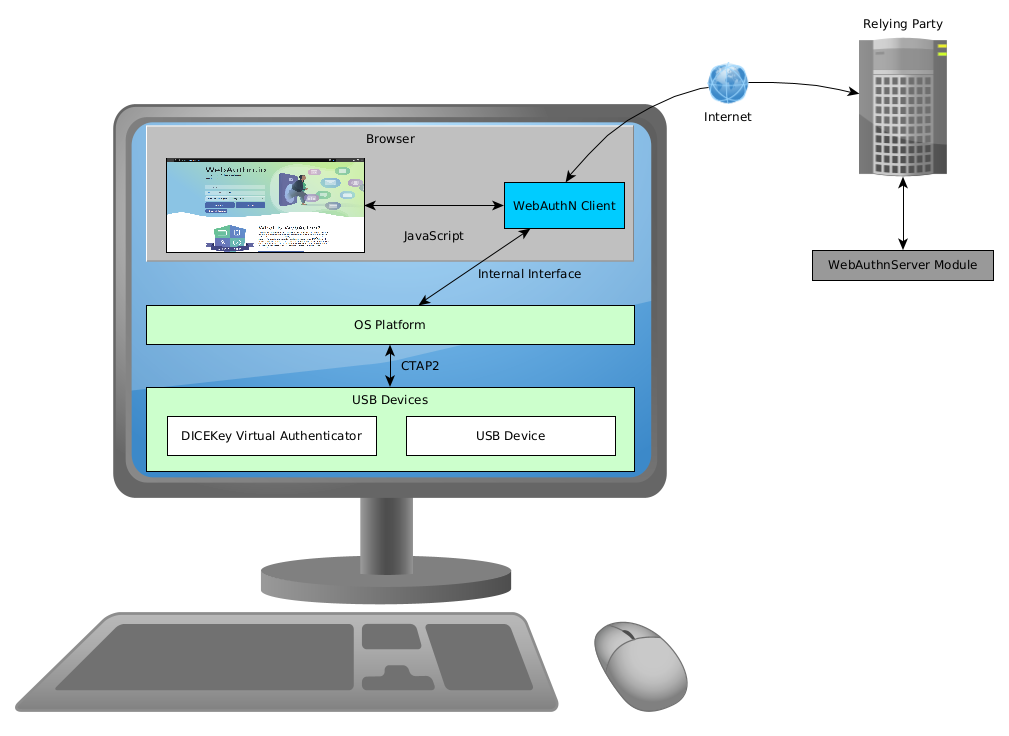}
\caption{WebAuthn and CTAP2 Overview}
\end{center}
\label{fig:webauthn_overview}
\end{figure}

Figure \ref{fig:webauthn_overview} provides an overview of how these different components fit together and the communication between them. The part we are most interested in for this proof of concept is the CTAP2 interface to the authenticator.

\subsection{History}
WebAuthn can be considered to be an output of the FIDO Alliance and their standardisation of firstly second factor tokens and subsequently roaming authenticators. The ancestry of both WebAuthn and CTAP2 can be traced back to the original U2F standard\footnote{\url{https://fidoalliance.org/specs/fido-u2f-v1.0-ps-20141009/fido-u2f-v1.0-ps-20141009-README.txt}}, initially written by Google and Yubico and subsequently managed by the FIDO alliance.

\subsubsection{CTAP 1}
CTAP1 was originally referred to as the U2F 1.0 protocol, released in 2014. It defined a standard for the provision of a Universal 2nd Factor (U2F). This provided a cryptographic challenge response protocol that provided assurance that the user was in possession of the 2nd factor. 

A further revision of CTAP 1 (U2F 1.2)\footnote{\url{https://fidoalliance.org/specs/fido-u2f-v1.2-ps-20170411/fido-u2f-README-v1.2-ps-20170411.txt}} was released in 2017 that further defined and standardised the communication with the hardware device. 

Through a number of iterations, U2F 1.0 eventually became the WebAuthn standard, whilst U2F 1.2 became the CTAP2 standard, with the original U2F functionality renamed to CTAP1.

This history has played a part in shaping the standard that exists today. The existence of hardware tokens predates the WebAuthn standard and as a result there are certain aspects of the standardisation that appear to be the result of standardising an existing device, rather than defining the standard from abstract concepts. The standard is an evolving one, so one would expect further refinement over the forthcoming years, with considerable developments already appearing the draft versions of the future standard.

Due to the rapidly changing nature of the living standard, this document is written with regards to the W3C Level 1 WebAuthn standard\footnote{\url{https://www.w3.org/TR/webauthn/}} and Client to Authenticator Protocol (CTAP) Proposed Standard, January 30, 2019. \footnote{\url{https://fidoalliance.org/specs/fido-v2.0-ps-20190130/fido-client-to-Authenticator-protocol-v2.0-ps-20190130.html}}

\subsection{CTAP2 Overview}
The CTAP2 standard\cite{ctap2} can be considered as covering two different aspects, one is the communication between a WebAuthn client and an authenticator, and the other is the processing of WebAuthn messages by the authenticator. The former is further broken down into USB\footnote{\url{https://fidoalliance.org/specs/fido-v2.0-ps-20190130/fido-client-to-authenticator-protocol-v2.0-ps-20190130.html\#usb}}, NFC\footnote{\url{https://fidoalliance.org/specs/fido-v2.0-ps-20190130/fido-client-to-authenticator-protocol-v2.0-ps-20190130.html\#nfc}} and Bluetooth\footnote{\url{https://fidoalliance.org/specs/fido-v2.0-ps-20190130/fido-client-to-authenticator-protocol-v2.0-ps-20190130.html\#bluetooth}} interfaces, although we are only concerned with the USB interface in this report. The latter describes the steps involved in responding to WebAuthn requests. 

\subsubsection{USB HID Interface}
The USB Human Interface Device (HID) interface section\footnote{\url{https://fidoalliance.org/specs/fido-v2.0-ps-20190130/fido-client-to-authenticator-protocol-v2.0-ps-20190130.html\#usb}} defines the packets, messages and transaction structure that is to be used for the USB HID interface. The fundamental structure involves three parts: packets, messages and transactions. Packets are the lowest level objects and consist of 64-byte objects that are transmitted via USB. They are always 64-bytes in length, with padding added as is necessary. There are two types of packet, \textit{initialization} packets and \textit{continuation} packets. \textit{Initialization} packets are always sent, and are the first packet of every message. If the payload does not fit into a single \textit{initialization} packet then $n$ \textit{continuation} packets are sent. One \textit{initialization} packet and 0 or more \textit{continuation} packets are concatenated to form a message.

Messages are either request or response messages. Requests are sent from the client to the authenticator and responses are sent in reply. Requests and responses are paired as a transaction. Each request should result in a response, and a response should never be sent without a request. Note, however, this may not be strictly true, if an error occurs during the processing of a packet, prior to the complete receipt of a request, then strictly speaking the error response will be sent prior to the complete request having been received or processed. In our opinion it would be better to state that the authenticator will never initiate communication and will only ever respond. Also note, keep-alive messages are sent during the processing of a request, but do not count as responses. In our implementation we wrap everything as a transaction for simplicity. 

Only a single transaction may be in operation at any one time. As such, once a request is initiated no further requests can be processed until a response has been sent. However, this is also not strictly true, since cancel messages can be sent outside the transaction constraints.

Each client must first initialise a channel by making a request on the broadcast channel. This will result in the authenticator issuing a unique Channel ID (CID) for that client to use for subsequent requests. As such, multiple clients can be concurrently supported by the authenticator but transactions, and therefore request processing, is not processed concurrently.

There are 7 mandatory CTAP2 HID commands as described below: 

\begin{description}
    \item [CTAPHID\_INIT] used to initialize a channel and return a new channel ID. Can also be used to re-initialize an existing channel.
    \item [CTAPHID\_CBOR] used to send WebAuthn messages, encoded as CBOR\footnote{\url{https://cbor.io/}}. Concise Binary Object Representation (CBOR) is the chosen encoding format for the WebAuthn messages. 
    \item [CTAPHID\_PING] a basic echo message for debugging and performance testing
    \item [CTAPHID\_CANCEL] cancels any ongoing message processing on the current Channel ID. Note this operates outside the transaction restrictions.
    \item [CTAPHID\_ERROR] only used in responses to signify an error has occurred.
    \item [CTAPHID\_KEEPALIVE] operate outside the transaction constraints to notify the client to the progress of ongoing processing.
    \item [CTAPHID\_MSG] encapsulates a CTAP1/U2F message in the U2F Raw Message Format. This is only implemented if supporting CTAP1/U2F messages, which is outside the scope of this project and report.
\end{description}

The CTAPHID\_MSG has been implemented at the USB level in our base framework but no message handling code has been implemented. Currently it will respond with an error message if such messages are received. This is because it is only used for backwards compatibility with U2F protocols, and is not required when evaluating the WebAuthn protocol messages that are sent via the CTAPHID\_CBOR messages, and as such was not within the scope of this project. Should such functionality be required it could be added by extending the process\_msg\_request function in ctap.py and adding the necessary functions to the authenticator framework. 

The messages described above do not map onto the WebAuthn standard, instead, WebAuthn messages are encoded as CBOR messages. Those messages are defined in the Authenticator Model in Section 5\footnote{\url{https://fidoalliance.org/specs/fido-v2.0-ps-20190130/fido-client-to-authenticator-protocol-v2.0-ps-20190130.html\#authenticator-api}} of the standard. However, there is not a clean split between WebAuthn messages and CTAP2 messages between these two parts of the standard. CTAP2 only messages also appear in Section 5 of the standard, for example, authenticatorGetClientPIN\footnote{\url{https://fidoalliance.org/specs/fido-v2.0-ps-20190130/fido-client-to-authenticator-protocol-v2.0-ps-20190130.html\#authenticatorClientPIN}} and authenticatorReset\footnote{\url{https://fidoalliance.org/specs/fido-v2.0-ps-20190130/fido-client-to-authenticator-protocol-v2.0-ps-20190130.html\#authenticatorReset}} standard. 

\subsubsection{Authenticator API}
The Authenticator API in Section 5\footnote{\url{https://fidoalliance.org/specs/fido-v2.0-ps-20190130/fido-client-to-authenticator-protocol-v2.0-ps-20190130.html\#authenticator-api}} of the CTAP2 standard defines the WebAuthn messages and the additional CTAP2 messages. The messages defined are as follows:

\begin{description}
    \item [authenticatorMakeCredential] maps to the WebAuthn makeCredential definition. Used to create a new PublicKeyCredential Source.
    \item [authenticatorGetAssertion]  partially maps to the GetAssertion definition in WebAuthn. Used to get an assertion of a PublicKeyCredential Source.
    \item [authenticatorGetNextAssertion] CTAP2 only, allows iterating through assertions for situations with multiple matching credentials.
    \item [authenticatorGetInfo] CTAP2 only, provides a list of capabilities to the client for the authenticator. Used by clients to construct a list of matching authenticators.
    \item [authenticatorClientPIN] CTAP2 only, defines how to get, set and check the Client PIN. Contains a series of sub commands for the different operations.
    \item [authenticatorReset] mentioned in WebAuthn, but not defined. Allows an authenticator to be reset.
\end{description}

The mapping between these definitions and WebAuthn is not exact, because of the presence of functionality such as the client PIN, which is not defined in WebAuthn, but appears in addition to the WebAuthn content in messages sent to the authenticator via CTAP2.

\section{A Virtual Authenticator}\label{sec:virtual_authenticator}
WebAuthn and CTAP2 are substantial standards dependent on multiple different implementations and entities. At the very least, a RP, a client, and an authenticator. Our objective was to build a virtual authenticator that would appear as a USB HID and therefore be capable of interacting with a client/browser as if it was a hardware authenticator. This would allow us not only to build a framework for further prototyping, but also examine the low-level aspects of the CTAP2 protocol.

\subsection{System Components}
In order to create a virtual authenticator a number of system levels services and components need to be used. In particular, the Linux USB Gadget system\cite{gadgetfs} is used to create a virtual device that will appear as a hardware authenticator to a WebAuthn client. 

\subsubsection{USB Gadget}
The USB Gadget system in Linux allows the creation of user-space USB devices \cite{gadgetfs}. Bottomley described using the Gadget system to create an authenticator~\cite{James_Bottomley}, and this work would later be included in the Solo Keys\footnote{\url{https://solokeys.com/}} project to facilitate their development simulator\footnote{\url{https://github.com/solokeys/solo-python}}. We modified the approach from Solo Keys to generate our USB Gadget, which we named DICEKey. In effect, this provides the end point descriptions as detailed in Section 8.1.8 of the CTAP2 standard \footnote{\url{https://fidoalliance.org/specs/fido-v2.0-ps-20190130/fido-client-to-authenticator-protocol-v2.0-ps-20190130.html\#usb-hid-implementation}}. Once created, a device will be appear in /dev/ called dicekey which we can then interface with using standard IO components within Python. 

The generated device is an HID, which is a particular standard for USB, more typically found in keyboards and mice. The standard allows a device to be used without special drivers. The built in device driver provides a standard HID packet, allowing easy reading and writing of the specified packet to allow communication with the device.

\paragraph{udev Rules} The permissions and naming of the device are set using udev rules. A sample rules files is provided with instructions of where to set those rules. It is necessary to set the udev rule prior to creating the device. The particular device is identified by a \textit{vendorId}. Originally this was the \textit{vendorId} provided in the Solo keys sample code. However, reliability problems were found when using that \textit{vendorId}. It appears that the same \textit{vendorId} is used for a device that allows firmware updates. On some Linux installs there is a service that looks for that specific \textit{vendorId} and captures the device to check for any possible firmware updates. This process fails due to it not being the same device, but the lock on the device is not released. To counter this an alternative \textit{vendorId} was used. The \textit{vendorId} used is allocated for open source projects\footnote{\url{https://github.com/obdev/v-usb/blob/master/usbdrv/USB-IDs-for-free.txt}}, and as such is shared by many open source and open hardware projects. \textit{vendorId}'s are not freely available and no official mechanism is provided for open source or open hardware projects. For development purposes the approach taken will work fine, but may struggle if deployed in the real-world due to comflicts with other devices.

\subsection{Python Packages and Modules}
The framework has been written in Python3 and consists of the following packages:

\begin{enumerate}
    \item hid
    \item ctap
    \item authenticator
    \item crypto
\end{enumerate}

\subsubsection{hid}
The hid package contains classes to read and write from the underlying USB device and to encode and decode the messages being sent and received.

\paragraph{usb.py}
The first part of the Python implementation handles the interface to the USB Gadget. In essence it opens a reader and a writer for the \textit{/dev/dicekey} device. It starts two threads, one to read from the device and one to write to it. The read thread loops forever attempting to read 64 bytes from the stream - where 64 bytes is the standardised size of all HID packets. 

The interface provides a listener structure to allow another class to register as the listener and be notified whenever a HID Packet arrives. Currently this is limited to a single listener, but it was designed with the intention to allow potentially multiple listeners, i.e. multiple authenticators in the future. The listener interface is defined in \textit{listener.py} and consists of just two method signatures \textit{received\_packet} and \textit{response\_sent}. \textit{received\_packet} will be called whenever an HIDPacket is received, it is the job of the listener to reassemble those packets into a complete message and transaction. \textit{response\_sent} is called once a response and been written to the channel, this notifies the listener to reset their transaction so they are ready to receive further incoming messages.

Once a packet is read it is loaded into a HIDPacket object, defined in \textit{packets.py}, and passed onto the registered listener for processing.

\paragraph{packets.py}
Defines the HID packet structures. There is an overarching HIDPacket which is subclassed by HIDInitializationPacket and HIDContinuationPacket. These two types of packet are defined in the CTAP2 standard. Due to the length limit of 64 bytes, if a message requires more space it will be split across one initialisation packet, and $n$ continuation packets. All messages will consist of exactly one HIDInitializationPacket and potentially many HIDContinuationPackets. From a developer perspective this is hidden, since the packet will only be passed on for processing once it is complete, i.e. all continuation packets have been received and concatenated. 

\paragraph{ctap.py}
Implements a listener for processing CTAP2 messages and making the appropriate calls to the authenticator class. It manages the transactions, as well as the reassembly of the incoming HIDPackets. Any messages that can be responded to independently of the authenticator is also implemented in this class, for example, ping requests that just echo the received message back. Additionally, it is the job the listener to enforce the transaction restrictions defined in the CTAP2 standard, effectively only allowing a single transaction to be in progress at any one time. The listener itself may operate outside this restriction in order to send error message rejecting an incoming request. Note however, that this does not run on an independent thread, so if the authenticator is currently processing a transaction and building a response, the thread will be occupied and any further message will be queued. However, if an HIDInitializationPacket has been received and a transaction started, but not executed because not all continuation packets have been received, error messages to other channels could be sent. This could be expanded to handle further concurrency, but since the standard itself currently does not support concurrent processing  implementing such  functionality has not been included.

For each CTAP2 defined message there will be an equivalent \textit{process\_MSG} function. Where \textit{MSG} represents the underlying message type. For example, \textit{process\_cbor\_request} or \textit{process\_init\_request}. The class implements a \textit{process\_msg\_request} which is intended to handle CTAP1/U2F messages, but simply responds with an error message as the authenticator does not implement CTAP1/U2F. Should future development wish to handle such message this function should be modified to call the relevant functions in the authenticator.

\subsubsection{ctap}
The ctap package contains the data structures, constants and encoding/decoding functionality necessary for interpreting the underlying CBOR messages that are sent and received as part of CTAP2. This is purely for message and transaction handling, not for processing the actual contents of the messages.

\paragraph{messages.py}Defines an object for each different type of message. These objects will contain various data handling functions, for example, encoding/decoding, and verification of message contents. The module contains the following classes:
\begin{itemize}
    \item CTAPHIDCMD
    \item CTAPHIDMsgRequest
    \item CTAPHIDMsgResponse
    \item CTAPHIDCancelRequest
    \item CTAPHIDCancelResponse
    \item CTAPHIDKeepAliveResponse
    \item CTAPHIDErrorResponse
    \item CTAPHIDWinkRequest
    \item CTAPHIDWinkResponse
    \item CTAPHIDPingRequest
    \item CTAPHIDPingResponse
    \item CTAPHIDCBORRequest
    \item CTAPHIDCBORResponse
    \item CTAPHIDInitRequest
    \item CTAPHIDInitResponse
\end{itemize}

\textbf{CTAPHIDCMD} is a superclass for all the other message classes and defines core functionality shared between other messages. For example, getting the channel ID, checking if it is complete or is awaiting for further continuation packets, and extracting the complete message payload - the data across all initialization and continuation packets. It also defines an abstract \textit{verify} method that is called on all received requests to verify the contents of the message are valid CBOR. It is the responsibility of the subclass to implement the appropriate verification logic in the subclass. Currently the \textit{verify} method is only implemented for incoming requests, responses constructed by the authenticator are assumed to be valid and not verified on construction.

There is also a static \textit{create\_message} function that takes an HIDPacket and constructs an appropriate message type based on the \textit{CMD} value in the HIDPacket. This method should be called to get an appropriate message type for onward processing. If additional messages are to be supported the \textit{create\_message} method should be expanded to construct an appropriately typed message.

\begin{figure}[t]
\includegraphics[width=\textwidth]{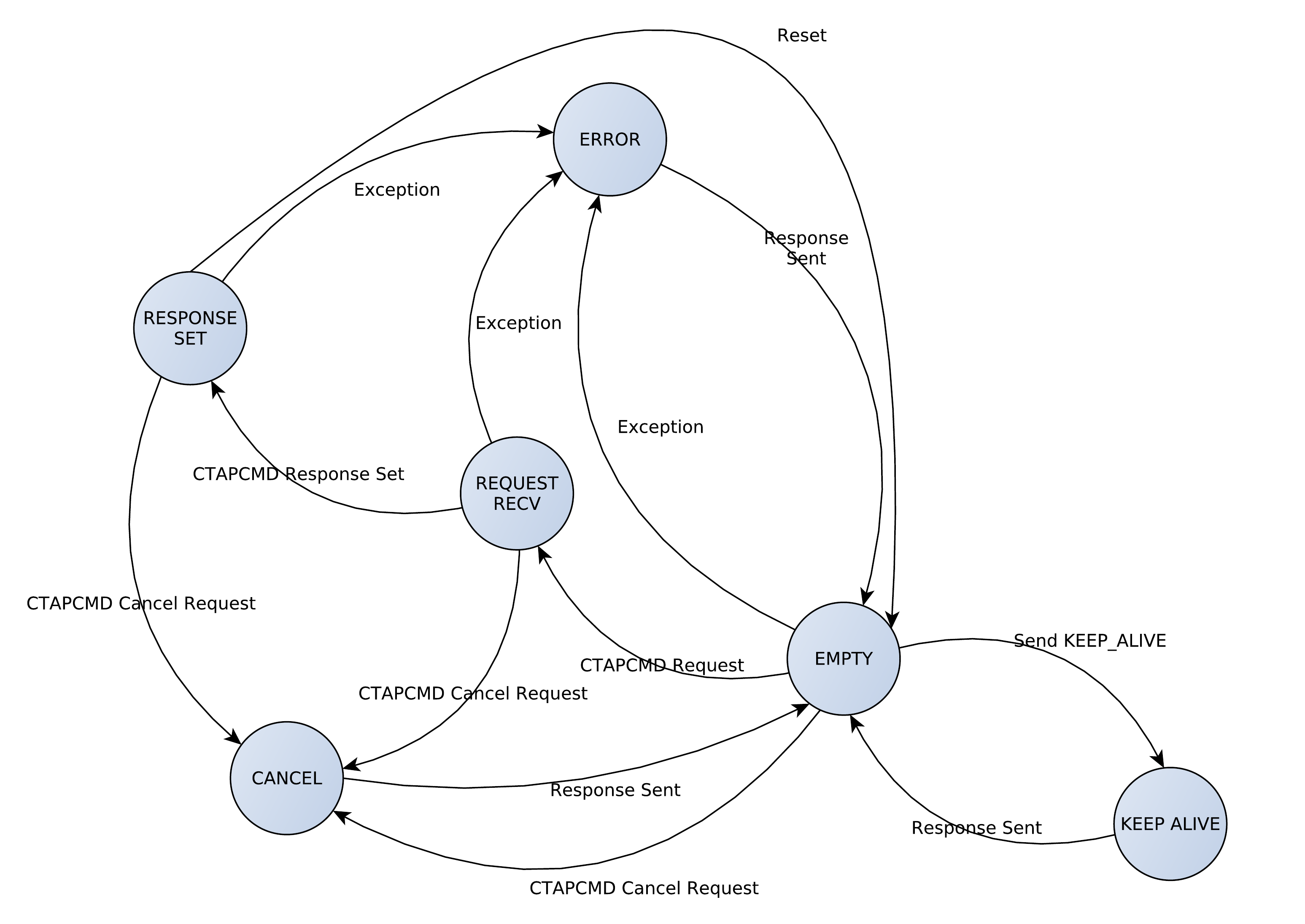}
\caption{Finite State Machine for Transactions}
\label{fig:fsm_transaction}
\end{figure}

\paragraph{transaction.py} contains a CTAPHIDTransaction class that manages the transaction state. A transaction is in effect a finite state machine as shown in Figure \ref{fig:fsm_transaction}, consisting of the following states:
\begin{itemize}
    \item EMPTY
    \item REQUEST\_RECV
    \item RESPONSE\_SET
    \item KEEP\_ALIVE
    \item CANCEL
    \item ERROR
\end{itemize}

An EMPTY state is the starting point for a transaction, and the state that is set when a reset occurs. In an EMPTY state a transaction can receive a request. Once the request is received the state changes to REQUEST\_RECV. Once the request is processed and a response set, the transaction moves to the RESPONSE\_SET state. Note, this only indicates the response has been set in the transaction, not that it has actually been sent. Once the response has been sent the transaction is reset back to EMPTY. 

There are also some additional states that can occur for potentially out-of-band messages, for examples, error messages that are sent irrespective of the state of a transaction. These messages operate outside the normal Request-Response flow, however, they are wrapped as transactions to aid simpler processing. As such, an Error message might consist of a single Error response, with no accompanying request. Similarly, the CANCEL state can be applied at any point. KEEP\_ALIVE messages are sent in an independent thread in order to keep the channel alive whilst some processing or user interaction is achieved. As such, there are response only messages without an accompanying request.

\paragraph{credential\_source.py}
Defines the \textit{PublicKeyCredentialSource} class that holds the details of a \textit{PublicKeyCredentialSource}. This class encodes and decodes the \textit{PublicKeyCredentialSource} as bytes to allow storage and retrieval. Additionally it provides functions for retrieving parts of the credential source, for example, the Relying Party Entity or the User Entity. It also provides access to the cryptographic material via calls to get the private key and the public key. Note, these methods are implemented using references to an abstract class, for example, \textit{AuthenticatorCryptoPrivateKey}, to allow it to be agnostic of the underlying crypto algorithm, which is managed by the associated crypto provider. It is the responsibility of the concrete implementation of \textit{AuthenticatorCryptoPrivateKey} and the associated other classes to allow the key to be serialized and deserialized appropriately for storage, for example, via the \textit{get\_encoded()} function. 

\paragraph{constants.py}
Contains a series of Enum classes that contain the constants used to define various parameters and values. Due to the variety of encoding methods, including CBOR and strings, it is easier to maintain those constants as Enums and provide a more informative variable name in the main code. This also hides the details of whether a field is referred to by an integer, a byte, or a string, allowing easier development.

\paragraph{exceptions.py}
Defines a series of base exceptions that are used to indicate an error. Rather than explicitly creating an error message and sending it, exceptions are used to raise the error and then have it caught higher up the stack. At that higher level the appropriate CTAP2 error message is constructed and sent to the client.

\paragraph{attestation.py}
Defines the structure for attestation statements. Currently this only contains a packed self-attestation. When further attestation formats and methods are implemented they should be added to this class to provide a single point for attestation construction.

\paragraph{keep\_alive.py}
Creates a new thread for a transaction that can be used to send keep alive messages at a pre-determined interval. A \textit{CTAPHIDKeepAlive} object is always created for each transaction, and passed on to the authenticator. If the authenticator wishes it can start the keep-alive sending, which will run in a new thread and continue until stopped or it expires. This runs independently of the transaction and sends keep-alive messages outside the normal request-response transaction. 

\subsubsection{ctap}
The crypto package contains both abstract definitions of a crypto provider and concrete implementations. The concept is to create an abstract definition of what a crypto provider needs to implement, for example, key pair generation and signing. When instantiated an authenticator loads crypto providers for various different algorithms, indexed by their registered COSE algorithm ID\footnote{\url{https://tools.ietf.org/html/draft-ietf-cose-webauthn-algorithms-04}}. That list of providers is used to advertise the available algorithms and match incoming requests to the appropriate crypto provider. From the perspective of the authenticator it is agnostic of the underlying crypto provider, allowing different implementations to be introduced and further providers be implemented without any changes to the underling authenticator.

\paragraph{algs.py}
Provides an enumerator of all the current COSE Algorithm identifiers. It is provided as a convenience to allow easier referencing to algorithm names, instead of the integer indexes that are not particularly informative.

\paragraph{crypto\_provider.py}Defines four abstract classes that define the required functionality of a crypto provider. The four classes are:
\begin{itemize}
    \item AuthenticatorCryptoPublicKey
    \item AuthenticatorCryptoPrivateKey
    \item AuthenticatorCryptoKeyPair
    \item AuthenticatorCryptoProvider
\end{itemize}

A \textit{AuthenticatorCryptoKeyPair} consists of a \textit{AuthenticatorCryptoPublicKey} and \textit{AuthenticatorCryptoPrivateKey}, which are self-explanatory in nature. The \textit{AuthenticatorCryptoProvider} is simple in nature, in that it defines a function to create a key pair and to load a key pair, as well as function to load just a public key from a COSE\footnote{\url{https://tools.ietf.org/id/draft-ietf-cose-rfc8152bis-struct-00.html}} encoded structure. The actual work of signing a message is performed in the \textit{AuthenticatorCryptoPrivateKey}.

\paragraph{es256\_crypto\_provider.py}Provides a concrete implementation of the ES256 crypto provider. As such, it contains concrete implementations of the four abstract classes:
\begin{itemize}
    \item ECCryptoKeyPair
    \item ECCryptoPrivateKey
    \item ECCryptoPublicKey
    \item ES256CryptoProvider
\end{itemize}

This implementation uses the SECP256R1 curve. There is a proposed standard to use the Koblitz curve\footnote{\url{https://tools.ietf.org/html/draft-jones-webauthn-secp256k1-00}}, but currently that is not supported. The underlying crypto operations are performed using the Python Cryptography library\footnote{url{https://cryptography.io/en/latest/}}, with keys encoded using PKCS8. There is currently an external dependency to the fido2 Python library\footnote{\url{https://pypi.org/project/fido2/}} in order to perform the COSE encoding of the public key. This is in practice an extremely simple undertaken, just arranging the values into an appropriate dictionary, and will in future be re-implemented within the codebase to remove the external dependency. 

\paragraph{tpm\_es256\_crypto\_provider.py} Provides an alternative ES256 crypto provider that is backed by a TPM. This demonstrates the flexibility of the approach, allowing the TPM variant to be swapped in for the Python cryptography variant with no change in the authenticator code. As such, it too implements the necessary classes:

\begin{itemize}
    \item TPMECCryptoKeyPair
    \item TPMECCryptoPrivateKey
    \item TPMECCryptoPublicKey
    \item TPMES256CryptoProvider
\end{itemize}
    
These classes act as a wrapper to the underlying TPM class, defined in \textit{/tpm/ibmtpm.py}. As such, the classes themselves may do little actual processing beyond encoding and decoding. For example, the \textit{TPMECCryptoPrivateKey} does not perform the signing itself, instead calling the TPM wrapper to load the relevant keys into the TPM and then perform the signing inside the TPM. 

\paragraph{./tpm/ibmtpm.py}\label{sec:ibmtpm} Defines the ctypes wrapper around the C TPM class. It is this class the manages the loading and unloading of keys into the TPM and the actual signing and key generation. The interface to the C TPM wrapper utilises a number of structures for passing data between Python and C. In particular, structures that define a byte\_array at the lowest level, and key structures at the higher level. For each C structure there is an equivalent Python class that handles the marshalling and unmarshalling of data types between Python and C. For example, there is \textit{RelyingPartyKey} structure defined in C, consisting of a key\_blob [KeyData] and a key\_point [KeyECCPoint]. There is a corresponding Python class called \textit{DICERelyingPartyKey} that allows loading the structure from the received C data as well as converting the underlying point into a Python EC Point for further processing.

\paragraph{credential\_wrapper.py} Defines an abstract class for performing credential wrapping. Credential wrapping occurs when a non-resident key is being created, i.e. a key that will reside outside of the accessible data stores of the authenticator, for example, on the relying party. The exact cryptographic algorithm to be used is left up to the authenticator. This is acceptable because only the authenticator needs to be able to decrypt and utilise the wrapped credential. However, by not standardising to a particular algorithm, there is a risk that authenticators could use weak encryption algorithms. In order to provide flexibility, and to potentially allow for future expansion, an abstract class was defined to provide the core functionality of key generation, wrapping and unwrapping. 

\paragraph{aes\_credential\_wrapper.py} Provides a concrete implementation using the AES256 key wrapping implementation provided by Python Cryptography, which uses an AES key wrapping algorithm with padding, as defined in RFC 5649. The key is generated as 32 random bytes and stored in the encrypted authenticator storage.

\subsubsection{Authenticator}
The Authenticator package contains the abstract and concrete classes necessary to define an authenticator. It is not a complete authenticator in itself, but provides the abstract class of an authenticator that must be implemented to build a working authenticator. However, it does provide concrete implementations of Authenticator storage classes as well as a UI implementation.

\paragraph{cbor.py} Defines the CBOR message response classes that are used to set the response parameters and then correctly encode them as CBOR, ready to be returned to the client. An overarching superclass, \textit{CBORRequest} is defined to provide core functionality, with specific message structures implemented in the respective subclasses:
\begin{itemize}
    \item GetClientPINResp
    \item MakeCredentialResp
    \item GetAssertionResp
    \item GetNextAssertionResp
    \item ResetResp
    \item GetInfoResp
\end{itemize}

\paragraph{datatypes.py} Defines various classes that manage access to data structures within requests, in particular various different parameter objects for the different requests. In addition to providing the necessary encoding and decoding functionality, they also provide a verify function to check for valid construction of the parameters. Additionally, an overarching exception class, DICEAuthenticatorException is defined. This exception class is used throughout the authenticator to raise exception that will subsequently be caught and returned as an appropriately formatted CTAP2 Error message. The various classes defined are as follows:

\begin{itemize}
    \item AuthenticatorVersion
    \item PublicKeyCredentialParameters
    \item PublicKeyCredentialRpEntity
    \item PublicKeyCredentialUserEntity
    \item AuthenticatorGetClientPINParameters
    \item PublicKeyCredentialDescriptor
    \item AuthenticatorGetAssertionParameters
    \item AuthenticatorMakeCredentialParameters
    \item DICEAuthenticatorException
\end{itemize}

\paragraph{preferences.py} Provides a minimal implementation of a JSON preferences file. Currently this is not linked to a UI and any preferences need to be manually set. A minimum number of preferences are currently supported, for example, whether the authenticator should default to storing credentials as resident keys or not, and the location of the authenticator storage file. This should be expanded to provide a greater range of preferences and integrated into the UI to provide easy access to the preferences. 

\paragraph{storage.py} Defines the abstract class for authenticator storage, \textit{DICEAuthenticatorStorage}. This provides the authenticator with a flexible framework for handling different storage mediums. The current concrete implementations are JSON based, but there is no reason the storage medium couldn't be a database, either local or remote. The objective is to define the abstract API and allow different storage mediums to be used as is desired. A number of abstract functions are defined, some obvious, for example, setting and getting the credential wrapper key. Others are more nuanced, for example, \textit{get\_credential\_source\_by\_rp} which is expected to provide a method for retrieving credentials related to particular Relying Party and filtering by an optionally provided allow\_list. There is also an \textit{DICEAuthenticatorStorageException} defined to handle storage exceptions. 

\paragraph{json\_storage.py} Defines two class to manage both plaintext and encrypted JSON storage. The \textit{JSONAuthenticatorStorage} defines a JSON based storage structure that writes all data to the file whenever a change occurs. Whilst this is not the most efficient approach, it avoids problems of caching data and having to gracefully shutdown the storage object. A set of storage keys are defined in the Enum STORAGE\_KEYS. The basic structure is a top-level JSONObject that has top level parameters, like global signature counter, AES wrapper key, client PIN object, etc. A ``credentials" field contains a JSONObject with JSONArrays indexed by Relying Party ID. The JSONArray associated with a particular Relying Party stores the credentials in the order they were generated, allowing the most recent to be prioritised, as is required by the CTAP2 standard. This approach is simpler than storing dates and then having to sort. 

The \textit{EncryptedJSONAuthenticatorStorage} class is a subclass of \textit{JSONAuthenticatorStorage} that provides an encrypted variant of the JSON storage. The subclass is extremely simple, just overriding the file reading and writing functions of \textit{JSONAuthenticatorStorage} to decrypt and encrypt respectively. The encryption key is derived from the user password that is associated with the authenticator. It would be possible to have distinct storage passwords and user verification passwords, but currently that is not the case. The underling encryption used is the Fernet authenticated encryption scheme defined by Python Cryptography\footnote{\url{https://cryptography.io/en/latest/fernet.html}}. Fernet uses AES128 in CBC mode with a random IV and PKCS7 padding. Authentication is provided via a SHA256 HMAC.

\paragraph{ui.py} Defines an abstract UI class and a concrete Command Line UI and a QT5 System Tray based UI. The abstract UI is defined in \textit{DICEAuthenticatorUI} with an accompanying \textit{DICEAuthenticatorListener} to receive UI events. The recommended UI requirements are fairly simple, requiring the following:
\begin{itemize}
    \item A user presence check
    \item A user verification check
    \item A password entry
    \item A shutdown method
\end{itemize}

Strictly speaking, only a shutdown method is absolutely necessary, however, some form of user presence check should also be implemented. The current Command Line UI is incomplete and only provides a shutdown function. The QT5 UI is the currently supported UI.

The QT5 UI is defined in \textit{QTAuthenticatorUI} which provides the concrete implementation of \textit{DICEAuthenticatorUI} and \textit{QTAuthenticatorUIApp} which provides the actual QT5 system tray application. The UI creates a system tray icon as well as a menu. Dialogs are superimposed to create the impression of notifications. Plain notifications are not used because on Linux they do not provide sufficient functionality to handle the user input requirement of User Verification and password entry. Additionally, the exact location of the system tray cannot be determined so the exact placement of the pseudo-notifications are independent of the location of the system tray icon - they will always appear in the top right hand corner.

Due to the way QT has to run on the main thread, it is necessary to split some of the initialisation functionality of an authenticator to after the UI has initialised to allow for retrieval of the user password, and the decryption of the authenticator storage. As such, there is a post\_ui loaded event that is fired once the UI has been created as is ready to receive requests. This is sent to the listeners so that they may continue their loading.

\paragraph{diceauthenticator.py} Contains the abstract definition of an authenticator. Whilst this class defines a number of abstract methods it also includes a large amount of generic processing to handle common processing that will occur across all authenticators. It provides the bridge between the various components, acting as the USBHID listener to receive incoming messages and setting itself (or its subclass) as the authenticator for the CTAPHID class. Examples of functionality that is generic is client PIN handling. The method for handling the client PIN is defined in the standard, including the crypto algorithms, and as such, does not vary from one authenticator to the next. As such, this functionality is fully implemented in the class. 

With regards to the client PIN it is worth noting that it uses its own instance of the ES256CryptoProvider in order to handle the crypto operations associated with the client PIN requests and responses. This is the defined cryptographic method in the standard and not open to variation. It does not reuse the ES256 crypto provider created as part of the crypto\_providers framework since it could be TPM based and not provide sufficient functionality to perform all PIN requests. It would be possible to expand the TPM functionality to include such functionality, but since the client PIN is entered into the client in any case, the security level is already reduced to that of software running on the machine, so probably does not warrant the high security properties of the TPM. 

In terms of WebAuthn and CTAP2 the most important function is \textit{process\_cbor} which handles incoming CBOR messages and calls the appropriate method for the different message types. Messages that require authentication processing, for example, making credentials, will call abstract methods, whilst requests that are generic will call concrete method within \textit{diceauthenticator.py}. Note, even though those concrete methods are provided, a developer could override them in the subclass should they wish to.

\subsubsection{dice\_key.py}
The \textit{dice\_key.py} is an example implementation of an authenticator using the above defined framework. This demonstrates how an authenticator can be implemented in a little over 500 lines of code using the above framework. This implementation uses a password to perform user verification, as well as the ES256 TPM backed crypto provider and an encrypted JSON storage.

One current limitation is the inability to have a Python based ES256 crypto provider alongside a TPM based ES256 crypto provider. To support side-by-side operation some form of priority management would be required, as well as additional fields stored in the credential to determine which crypto provider was used to create. Whilst it would be possible to develop, it is beyond the scope of this implementation. 

\subsubsection{Logging}
A logging structure has been implemented that provides a number of different logs to facilitate both analysis of the protocols components and overall debugging. The logging consists of the following, each of which is stored in a separate file and archived with a timestamp between runs. As such, each file will contain only the logging messages for a single instantiation of the code.

\paragraph{debug log} The debug log contains all log messages from all the different logs and can be thought of as the super log. In addition to being written to ``./logs/debug\_\textit{TIMESTAMP}" it is also written to System Out during the run of the application. 

\paragraph{usbhid log} Contains a log of received and sent USB HID messages. The raw contents of the messages is also logged. This log is formatted to show how messages are broken down into continuation packets and as well. Messages are logged in a JSON format to allow for possible automated analysis, the payload data is logged in Hex.

\paragraph{ctap log} Contains the log messages relating to the actual CTAP2 protocol. It will show the type of CTAP2 message received, for example, INIT or CBOR, as well as its contents. It will also show the transaction handling, in particular where an initialization packet is received and then continuation packets are added until the complete message has been received. Responses will also be logged as they are constructed and updated. In addition the state of the transaction is logged. If keep-alive messages are being sent these will also be very common in the log. 

\paragraph{auth log} The auth log contains logging messages from the authenticator itself. This will typically show the types of parameters and options selected and the outcomes of their processing. 

\paragraph{TPM Log} 
The TPM log can be found in ``./data/tpm/tpm\_log\_\textit{TIMESTAMP}". It is a dedicated log from the underlying TPM library code. Depending on the log level chosen (\textit{levels: 1 -- errors only, 2 -- basic information and 3 -- full information}), it will show the steps the TPM has taken in loading, creating and using keys as well as the unloading and cleaning up of the memory.

\input{Webauthn_tpm}

\subsection*{Acknowledgments}
This work has been supported by the following EPSRC projects: Improving customer experience while ensuring data privacy for intelligent mobility (EP/N028295/1) and an Impact Accelerator Project (IAA) project (EP/R511791/1).


\end{document}

%% file: Webauthn_tpm.tex
\section{TPM}
In addition to the development of the virtual authenticator, a further authenticator variant was implemented which used a TPM to create the necessary keys, use them for signing a \wan\ challenge and wrap them for storage outside of the TPM. Most of the software for the authenticator was written in Python, as described in the earlier sections, with a defined interface to the \cpp\ software used to access the TPM. 

This section describes how the TPM keys are organised, the software used to access the TPM  and the \cpp\ \textbf{--} Python interface.

\subsection{TPM Keys for the Authenticator}
The TPM is a small chip on the motherboard of a workstation used to provide secure cryptographic functions for the platform. The functions offered by this device are specified by the Trusted Computing Group~\cite{tcg.group}. Particularly relevant for this work is the ability to create and securely manage cryptographic keys. These TPM keys are stored in hierarchies (for example, see Figure~\ref{fig:key_hierarchy} which shows the key hierarchy used here). Apart from the top of the hierarchy, each key has a parent and this parent is used to wrap the key's sensitive data for storage outside of the TPM. For the key at the top of the hierarchy the sensitive data never leaves the TPM. These (primary) keys can be made persistent and stored inside the TPM, or re-generated as required. 

\input{Webauthn_tpm_keys}

The storage key hierarchy is used for the TPM keys for the authenticator. For this initial development the flat key hierarchy shown in Figure~\ref{fig:key_hierarchy} is used. The different levels are:
\paragraph{Storage root key} The storage root key is generated and made persistent, so it is always available. It is not password protected as it needs to be available to other users of the TPM.   
\paragraph{User key} User keys are storage keys, they are intermediate keys in the hierarchy used to protect their children. In the authenticator these keys are password protected, they could also be protected against dictionary attack, but that is not done here. The password is provided by the authenticator and not requested from the user except when they start to use the system.
\paragraph{Relying Party key} RP keys are signing keys used to sign the challenges received from their relying party. They can be password protected, but in the authenticator any password is automatically provided, the user is not asked to supply one.

\subsection{The \cpp\ TPM Class}
The interface to the TPM is provided by a \cpp\ class, \verb|Web_authn_tpm|, this takes care of all of the operations needed by the TPM authenticator and stores results of the operations for access by the Python software. To access the TPM, which can be a TPM simulator~\cite{ibmTPM}, or a hardware TPM, we needed to use a TCG Software Stack (TSS). There are several options available on Linux: the IBM TPM software stack~\cite{ibmTSS} and an open-source software (OSS) implementation of the TCG TSS standard~\cite{tpm2TSS, tss.spec}. In this project we used the IBM TSS as it is well supported and we had used it previously, when the other option was not so well developed. The TPM class provides the following calls:
\paragraph{setup} Setup the \verb|Web_authn_tpm| class. If it is not already in place, it creates the storage root key (SRK) and makes it persistent.
\paragraph{set\_log\_level} Sets the log level to be used. Options are: 1 -- errors only, 2 -- basic information and 3 -- full information (for debugging).
\paragraph{create\_and\_load\_user\_key} Creates a new user (storage) key and loads it ready for use. If a user key is already loaded, it and its relying party key (if one is loaded) are flushed and their data removed.
\paragraph{load\_user\_key} Loads a user key ready for use. If a user key is already loaded, it and its relying party key (if one is loaded) are flushed and their data removed.
\paragraph{create\_and\_load\_rp\_key} Creates and loads a relying party key. If a relying party key is already loaded, it is flushed and its data removed.
\paragraph{load\_rp\_key} Loads a relying party key ready for use. If a different relying party key is already loaded it is flushed and its data removed.
\paragraph{sign\_using\_rp\_key} Signs a digest from the relying party and returns the signature.
\paragraph{flush\_data} Flushes the keys and removes their data.
\paragraph{get\_last\_error} Returns the last error and then resets the error string. Used to return an error when a command fails. As interfacing to Python, the class does not throw any exceptions. 


\subsection{Interfacing with Python}
Information is passed to-and-fro using structures containing \verb|Byte_array|s:
\begin{verbatim}
struct Byte_array
{
    uint16_t size{0};
    Byte* data{nullptr};
}
\end{verbatim}
This structure can be used by \cpp\ and Python. It is deliberately kept simple with the rule that whoever allocates the memory is also responsible for freeing it. The structures used  are:
\paragraph{Key\_data} The data returned from the TPM and consisting of the public and private data for the key.
\begin{verbatim}
struct Key_data
{
 Byte_array public_data;
 Byte_array private_data;
};
\end{verbatim}
\newpage
\paragraph{Key\_ecc\_point} The public ECC key, a point on the chosen ECC curve (NIST P\_256).
\begin{verbatim}
struct Key_ecc_point
{
  Byte_array x_coord;
  Byte_array y_coord;
};
\end{verbatim}
\paragraph{Relying\_party\_key} The relying party key data and the corresponding public key (ECC point).	
\begin{verbatim}
struct Relying_party_key
{
  Key_data key_blob;
  Key_ecc_point key_point;
}; 
\end{verbatim}
\paragraph{Ecdsa\_sig} The ECDSA signature returned from the TPM.
\begin{verbatim}
struct Ecdsa_sig
{
  Byte_array sig_r;
  Byte_array sig_s;
};
\end{verbatim}
The data to be signed is just passed in a \verb|Byte_array|. Note that it cannot be larger than the size of the hash being used (SHA256 in our case).

\subsubsection{The \cpp\ Python Interface}
The \cpp\ Python interface contains routines that map directly onto those provided by the \verb|Web_authn_tpm| class. All of the routines make use of the data structures described above and several standard data types (\verb|int32_t| and \verb|const char*|). The routines are:
\begin{verbatim}
extern "C" {
// Allocate memory for the TPM class and return a void* pointer
// to it 
void* install_tpm();

// Setup the TPM
int32_t setup_tpm(void* v_tpm_ptr, bool use_hw_tpm,
    const char* tpm_data_dir, const char* log_filename);

// Set the logging level
TPM_RC set_log_level(void *v_tpm_ptr, int log_level);

// Return the last error
const char* get_last_error(void* v_tpm_ptr);

// Call the TPM class' destructor, flushing any keys from the
// TPM and freeing any memory, set v_tpm_ptr=nullptr
void uninstall_tpm(void* v_tpm_ptr);

// No parent authorisation as we are using the SRK with no
// password.
Key_data create_and_load_user_key(void* v_tpm_ptr,
    Byte_array user, Byte_array key_auth);

// No parent authorisation as we are using the SRK with no
// password, key authorisation not needed to load the key.
TPM_RC load_user_key(void* v_tpm_ptr, Key_data kd,
    Byte_array user);

Relying_party_key create_and_load_rp_key(void* v_tpm_ptr,
    Byte_array relying_party, Byte_array user_auth,
    Byte_array rp_key_auth);

Key_ecc_point load_rp_key(void* v_tpm_ptr, Key_data kd,
    Byte_array relying_party, Byte_array user_auth);

Ecdsa_sig sign_using_rp_key(void* v_tpm_ptr,
    Byte_array relying_party, Byte_array signing_data,
    Byte_array rp_key_auth);

TPM_RC flush_data(void* v_tpm_ptr);

} // end of extern "C"
\end{verbatim} 
\noindent Note:
\begin{enumerate}
\item Where a routine returns a data structure the memory for this structure is allocated in the \cpp\ code and should also be freed there. Conversely, the data for parameters of these routines is allocated in the Python code and should be freed there.
\item The routines provided are defined \verb|extern "C"| so that their names are not `mangled' when the \cpp\ code is compiled.
\item The \verb|install_tpm()| routine allocates memory for the \verb|Web_authn_tpm| class, casts the pointer that is returned to a \verb|void*| and returns it. The Python code stores this pointer and it is passed to all of the interface routines when they are called. Each routine casts the pointer back to a \verb|Web_authn_tpm*| and this is then used to call the relevant class' member function. 
\item \verb|TPM_RC| is the TPM return code and is zero for a successful result. It is an alias for \verb|int32_t|.
\end{enumerate}
As an example of the use of this interface, here is the code for \verb|load_rp_key|:
\begin{verbatim}
Key_ecc_point load_rp_key(void* v_tpm_ptr, Key_data kd,
    Byte_array relying_party, Byte_array user_auth)
{
    if (v_tpm_ptr==nullptr) {
        return Key_ecc_point{{0,nullptr},{0,nullptr}};
    }
	
    Web_authn_tpm* tpm_ptr=
        reinterpret_cast<Web_authn_tpm*>(v_tpm_ptr);
	
    std::string rp_str=byte_array_to_string(relying_party);
    std::string user_auth_str=byte_array_to_string(user_auth);
	
    return tpm_ptr->load_rp_key(kd,rp_str,user_auth_str);
}.
\end{verbatim}

\subsubsection{The Python \cpp\ Interface}
Python provides a number of interfaces to the C, including Cython, ctypes, and writing a Python C Extension. The quickest approach is ctypes, which provides a wrapper around a dynamic C library and provides functionality to marshall and unmarshall various datatypes. In that regard it is similar in concept to Java JNI. The structure is fairly simple, firstly the library has to be loaded, which returns a Python object from which function calls within the library can be called. 

\begin{verbatim}
self._tpm = ctypes.cdll.LoadLibrary(PATH_TO_LIBRARY)
\end{verbatim}

The next step is to define the parameter types and return types for the functions that are going to be called. ctypes uses this information to correctly instantiate appropriate Python objects from C memory locations. As well as the primitive datatypes already implemented by ctypes it is possible to implement custom objects by subclassing primitive ctypes.

For example, if we consider the ByteArray structure, we can create the equivalent structure in Python using the following:

\begin{verbatim}
class ByteArray(ctypes.Structure):
    _fields_ = [('size', ctypes.c_uint16),
                ('data', ctypes.POINTER(ctypes.c_byte))]
\end{verbatim}

This closely matches the equivalent C definition, as one would expect. It is also possible to nest these definitions in the same way a C. For example, the KeyECCPoint structure refers to two ByteArray structures:

\begin{verbatim}
class KeyECCPoint(ctypes.Structure):
    _fields_ = [('x_coord', ByteArray),
                ('y_coord', ByteArray)]
\end{verbatim}

When defining the parameter types and return types we make reference to these structures and the provided primitive types implemented by ctypes. For example, the load\_user\_key method, as defined in the C interface can be defined in Python by including the following two lines:

\begin{verbatim}
self._tpm.load_user_key.restype = ctypes.c_int
self._tpm.load_user_key.argtypes = [ctypes.c_void_p,KeyData,ByteArrayStr]
\end{verbatim}

This indicates that the function returns a C int and that there are three parameters to the function, a C void pointer, a KeyData structure, and a ByteArrayStr structure.

Once defined the actual function can be called by invoking the load\_user\_key function:

\begin{verbatim}
username_byte_array =  ByteArrayStr(
    len(user_key.username),user_key.username.encode())
self._tpm.load_user_key(
    self._tpm_ptr, user_key.get_as_key_data_struct(), username_byte_array))
\end{verbatim}

The above shows us creating a new ByteArrayStr structure from the username, the calling the load\_user\_key function and passing the appropriate parameters. As described in Section \ref{sec:ibmtpm} we have also implemented wrapper classes to make converting between C and Python easier, hence the call to get\_as\_key\_data\_struct which returns the user\_key in the appropriate C types structure.
\subsection{Future Work}
If the keys are to be used from different platforms, the keys need to be migratable. This can be achieved by using the mechanism for key duplication defined in the TPM standard. The user keys can be made migratable and the relying party keys can then be migrated with them without any extra processing required (provided they are created with the correct properties). The protocol for key migration requires a password to protect the migrating key blob, this can be generated by the TPM, but how to securely store it for future use would need careful consideration.

For this initial demonstration a flat key hierarchy was used. At least one extra level of indirection could be used to make migration of all of the users keys more straightforward, either for backup, or for transfer to another TPM. So the user's keys could be children on a migratable Web\_authn root key and when this is migrated all of the other keys can move along with it.

The work so far assumes that we are just focussing on using platforms with TPMs on board, Another avenue to explore would be the use of other secure environments, Intel SGX, for example.



%% file: Webauthn_tpm_keys.tex

\newlength{\picturewidth}
\newlength{\pictureheight}
\newlength{\boxmargin}
\newlength{\boxsep}

\newcommand{\textinbox}[3]{	\put(#1,#2){\makebox(0,0){\framebox(2\boxmargin+\widthof{#3},2\boxmargin+\heightof{#3}){#3}}}
}

\setlength{\boxmargin}{3pt}
\setlength{\boxsep}{10pt}
\setlength{\picturewidth}{320pt}
\setlength{\pictureheight}{136pt}

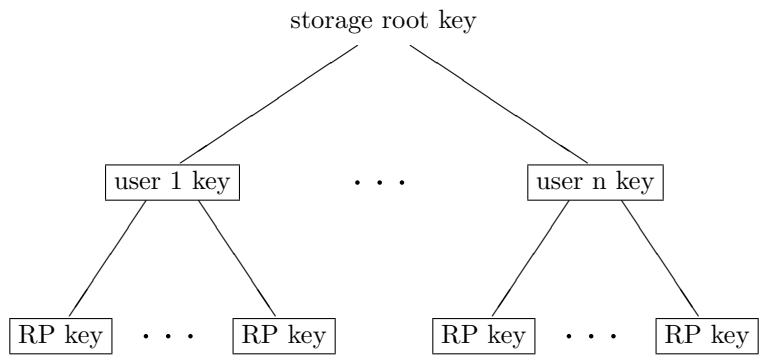
\begin{figure}[ht]
	\centering
	\unitlength=1pt
	\begin{picture}(\picturewidth,\pictureheight)
	\put(\picturewidth/2,\pictureheight-\heightof{gk}){\makebox(0,0){storage root key}}
	\textinbox{\picturewidth/4}{\pictureheight/2}{user 1 key}
	\put(\picturewidth/2-\boxsep,\pictureheight-16pt){\line(-3,-2){67}}
	\put(\picturewidth/2+\boxsep,\pictureheight-16pt){\line(3,-2){67}}
	\put(\picturewidth/2,\pictureheight/2){\makebox(0,0){\huge\ldots}}
	\textinbox{3\picturewidth/4}{\pictureheight/2}{user n key}
	\textinbox{\picturewidth/4-\widthof{RP key}-\boxsep}{\heightof{RP}+\boxmargin}{RP key}
	\put(\picturewidth/4,\heightof{RP}+\boxmargin){\makebox(0,0){\huge\ldots}}
	\textinbox{\picturewidth/4+\widthof{RP key}+\boxsep}{\heightof{RP}+\boxmargin}{RP key}
	\put(\picturewidth/4-\boxsep,\pictureheight/2-7pt){\line(-2,-3){29}}
	\put(\picturewidth/4+\boxsep,\pictureheight/2-7pt){\line(2,-3){29}}
	\textinbox{3\picturewidth/4-\widthof{RP key}-\boxsep}{\heightof{RP}+\boxmargin}{RP key}
	\put(3\picturewidth/4,\heightof{RP}+\boxmargin){\makebox(0,0){\huge\ldots}}
	\textinbox{3\picturewidth/4+\widthof{RP key}+\boxsep}{\heightof{RP}+\boxmargin}{RP key}
	\put(3\picturewidth/4-\boxsep,\pictureheight/2-7pt){\line(-2,-3){29}}
	\put(3\picturewidth/4+\boxsep,\pictureheight/2-7pt){\line(2,-3){29}}
	\end{picture}
	\caption{The TPM key hierarchy used in for the authenticator}
	\label{fig:key_hierarchy}
\end{figure}